\documentclass[aps,showpacs,preprint,footinbib,preprintnumbers]{revtex4}

\def\ltap{\ \raise.3ex\hbox{$<$\kern-.75em\lower1ex\hbox{$\sim$}}\ }
\def\gtap{\ \raise.3ex\hbox{$>$\kern-.75em\lower1ex\hbox{$\sim$}}\ }
\def\gl{\ \raise.4ex\hbox{$>$\kern-.75em\lower1ex\hbox{$<$}}\ }

\usepackage[normalem]{ulem}  
\usepackage[dvips]{color} 
\usepackage{amsmath,amssymb}
\usepackage{booktabs}
\usepackage{mathrsfs}
\usepackage[dvipdfmx]{graphicx}
\usepackage{graphicx}
\usepackage{multirow}
\usepackage{subfigure}

\renewcommand\sout{\bgroup \color{red} \ULdepth=-.5ex \ULset}

\def\ket#1{\mathinner{|{#1}\rangle}}
\def\braket#1{\mathinner{\langle{#1}\rangle}}

\begin{document}

\begin{flushright}
KEK-TH-1677 \\ J-PARC-TH-30
\end{flushright}

\title{Decays of $Z_b \rightarrow \Upsilon \pi$ via triangle diagrams in heavy meson molecules}
\author{S.~Ohkoda$^1$}
\author{S.~Yasui$^2$}
\author{A.~Hosaka$^{1,3}$}
\affiliation{$^1$Research Center for Nuclear Physics (RCNP), 
Osaka University, Ibaraki, Osaka, 567-0047, Japan}
\affiliation{$^2$KEK Theory Center, Institute of Particle and Nuclear
Studies, High Energy Accelerator Research Organization, 1-1, Oho,
Ibaraki, 305-0801, Japan}
\affiliation{$^3$J-PARC Branch, KEK Theory Center, Institute of Particle and Nuclear Studies,
KEK, Tokai, Ibaraki, 319-1106, Japan}

\begin{abstract}
 Bottomonium-like resonances $Z_b$(10610) and $Z_b^{\prime}(10650)$
 are good candidates of hadronic molecules composed of $B\bar{B}^{\ast}$ (or $B^{\ast}\bar{B}$)
 and $B^{\ast}\bar{B}^{\ast}$, respectively.
 Considering $Z_b^{(\prime)}$ as heavy meson molecules, we investigate the decays of $Z_b^{(\prime)+} \rightarrow \Upsilon (nS) \pi^+$ in terms of 
 the heavy meson effective theory.
 We find that the intermediate $B^{(\ast)}$ and $\bar{B}^{(\ast)}$ meson loops and the form factors play a 
 significant role to reproduce the experimental values of the decay widths.
 We also predict the decay widths of $Z_c^+ \rightarrow J/\psi \pi^+$ and $\psi(2S)\pi^+$ for a charmonium-like resonance $Z_c$ which has been reported recently in experiments.
\end{abstract}
\maketitle 

Two charged bottomonium-like resonances $Z_b(10610)$ and $Z_b^{\prime}(10650)$ were 
reported in the processes $\Upsilon(5S) \rightarrow \Upsilon(nS)\pi^{+}\pi^-$ ($n=1,2,3$) and
$\Upsilon(5S) \rightarrow h_b(mP)\pi^+ \pi^-$ ($m=1,2$) \cite{Collaboration:2011gja,Belle:2011aa}.
Their quantum numbers are $I^G(J^P) = 1^+(1^+)$, which indicates that the quark content of 
$Z_b^{(\prime)}$ must be four quarks as minimal constituents such as $\ket{b\bar{b}u\bar{d}}$.
The reported masses and decay widths of the two resonances are 
$M(Z_b(10610)) = 10607.4 \pm 2.0$ MeV,
$\Gamma(Z_b(10610)) = 18.4 \pm 2.4$ MeV and 
$M(Z_b(10650)) = 10652.2 \pm 1.5$ MeV,
$\Gamma(Z_b(10650)) = 11.5 \pm 2.2$ MeV,
showing that the masses are very close to the $B\bar{B}^{\ast}$ (or $B^{\ast}\bar{B}$) and $B^{\ast}\bar{B}^{\ast}$ thresholds, 
respectively.
In view of these facts, $Z_b$ and $Z_b^{\prime}$ are likely molecular states of 
two $B^{(\ast)}$ and $\bar{B}^{(\ast)}$ mesons~\cite{Bondar:2011ev,Ohkoda:2011vj,Ohkoda:2012rj}.

More recently, Belle reported the branching fractions of each channel in three-body
decays from $\Upsilon (5S)$ \cite{Adachi:2012cx}, the results of which are summarized in Table.~\ref{decayratio}.
They show a remarkable feature of $Z_b^{(\prime)}$.
One is that the dominant decay processes are channels to open flavor mesons, 
$\rm{Br}(Z_b^{+} \rightarrow B^{+}\bar{B}^{\ast 0} +B^{\ast +} \bar{B}^{0}) = 0.860$ and 
${\rm Br}(Z_b^{\prime +} \rightarrow B^{\ast +}\bar{B}^{\ast 0}) = 0.734$.
This is consistent with the naive consideration from the molecular picture.
Another point is in the ratios of the decay widths to a bottomonium and a pion,
where it is important to notice the following two facts.
Firstly, $h_b (mP) \pi^{+}$ decays are not suppressed in spite of their spin-flip processes 
of heavy quarks from $\Upsilon (5S)$.
In general, the spin-nonconserved decay in the strong interaction should be suppressed 
due to a large mass of $b$ quark.
Nevertheless, the spin-conserved decay $Z_b^{(\prime) +} \rightarrow \Upsilon(nS) \pi^+$ and 
spin-nonconserved one $Z_b^{(\prime) +} \rightarrow h_b(mP) \pi^{+}$ occur in comparable ratios.
The previous studies suggest that molecular picture explains well this behavior \cite{Bondar:2011ev,Ohkoda:2012rj}: if the $Z_b^{(\prime)}$ is a molecular state, 
the wave function is a mixture state of heavy quark spin singlet and triplet.
Then, $Z_b^{(\prime)}$ is possible to decay into both channels.
Secondly, the decay ratios are not simply proportional to the magnitudes of the phase space.
In particular, the branching fraction of $Z_b^{(\prime) +} \rightarrow \Upsilon(nS) \pi^+$ is only 
approximately ten percents of the one of $Z_b^{(\prime) +} \rightarrow \Upsilon(2S) \pi^+$ although 
the phase space of $\Upsilon(1S) \pi^+$ is larger than the one of $\Upsilon(2S) \pi^+$.
In fact, $\Gamma (Z_b^{(\prime) +} \rightarrow \Upsilon(3S) \pi^+)$ is approximately half a size of 
$\Gamma (Z_b^{(\prime) +} \rightarrow \Upsilon(2S) \pi^+)$, which is still wider than the 
$\Gamma (Z_b^{(\prime) +} \rightarrow \Upsilon(1S) \pi^+)$.
The mechanism of this behavior is not still elucidated completely
and needs detailed considerations. 
In this paper, we focus on the strong decays $Z_b^{(\prime) +} \rightarrow \Upsilon(nS)\pi^+$ and 
analyze their decay widths as hadronic molecules.
This study will also provide a perspective for the internal structure of $Z_b^{(\prime)}$.
Our approach also applies to the decays of $Z_c(3900)$, which is charged charmonium-like resonance reported both by the BESIII Collaboration~\cite{Ablikim:2013mio} and by
the Belle collaboration~\cite{Liu:2013dau}.

\begin{table}[tbp]
\caption{Branching ratios (Br) of various decay channels from $Z_b(10610)$ and $Z_b^{\prime}(10650)$.}
\begin{tabular}{c|c|c}
 \hline
 channel & Br of $Z_b$ & Br of $Z_b^{\prime}$,  \\
 \hline
 $\Upsilon(1S) \pi^+$  & $0.32 \pm 0.09$ & $0.24 \pm 0.07$ \\
 $\Upsilon(2S) \pi^+$  & $4.38 \pm 1.21$ & $2.40 \pm 0.63$ \\ 
 $\Upsilon(3S) \pi^+$  & $2.15 \pm 0.56$ & $1.64 \pm 0.40$ \\
 $h_b(1P) \pi^+$ & $2.81 \pm 1.10$ & $7.43 \pm 2.70$ \\
 $h_b(2P) \pi^+$ & $2.15 \pm 0.56$ & $14.8 \pm 6.22$ \\
 $B^{+}\bar{B}^{\ast 0} + B^{\ast +}\bar{B}^{0}$ & $86.0 \pm 3.6$ & $-$ \\
 $B^{\ast +}\bar{B}^{\ast 0}$ & $-$ & $73.4 \pm 7.0$
\end{tabular} 
\label{decayratio}
\end{table}
%

\

%
\begin{figure}[tbp]
 \subfigure{\ 
 \includegraphics[clip,width=50mm]{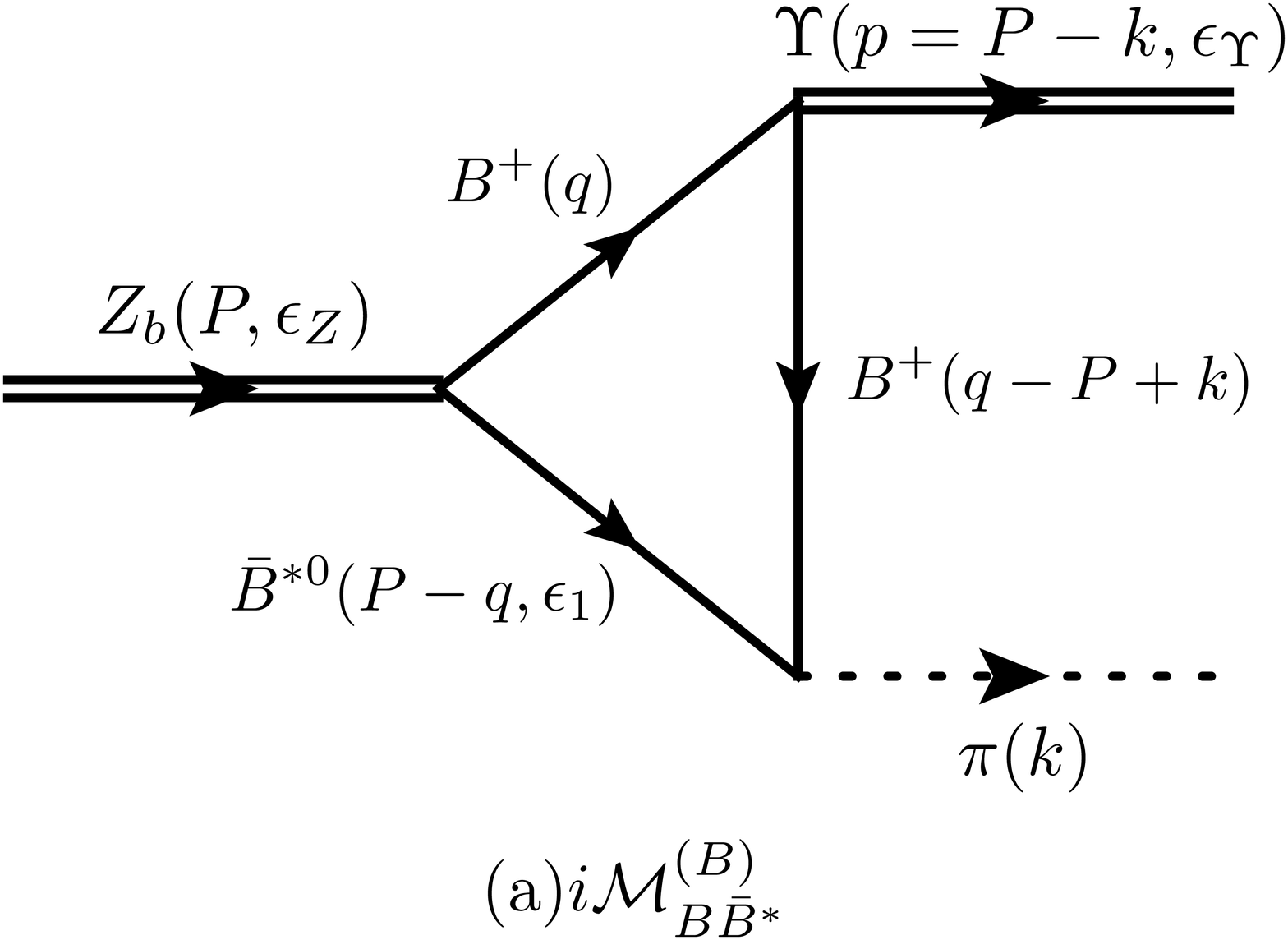} }\
 \subfigure{\
 \includegraphics[clip,width=50mm]{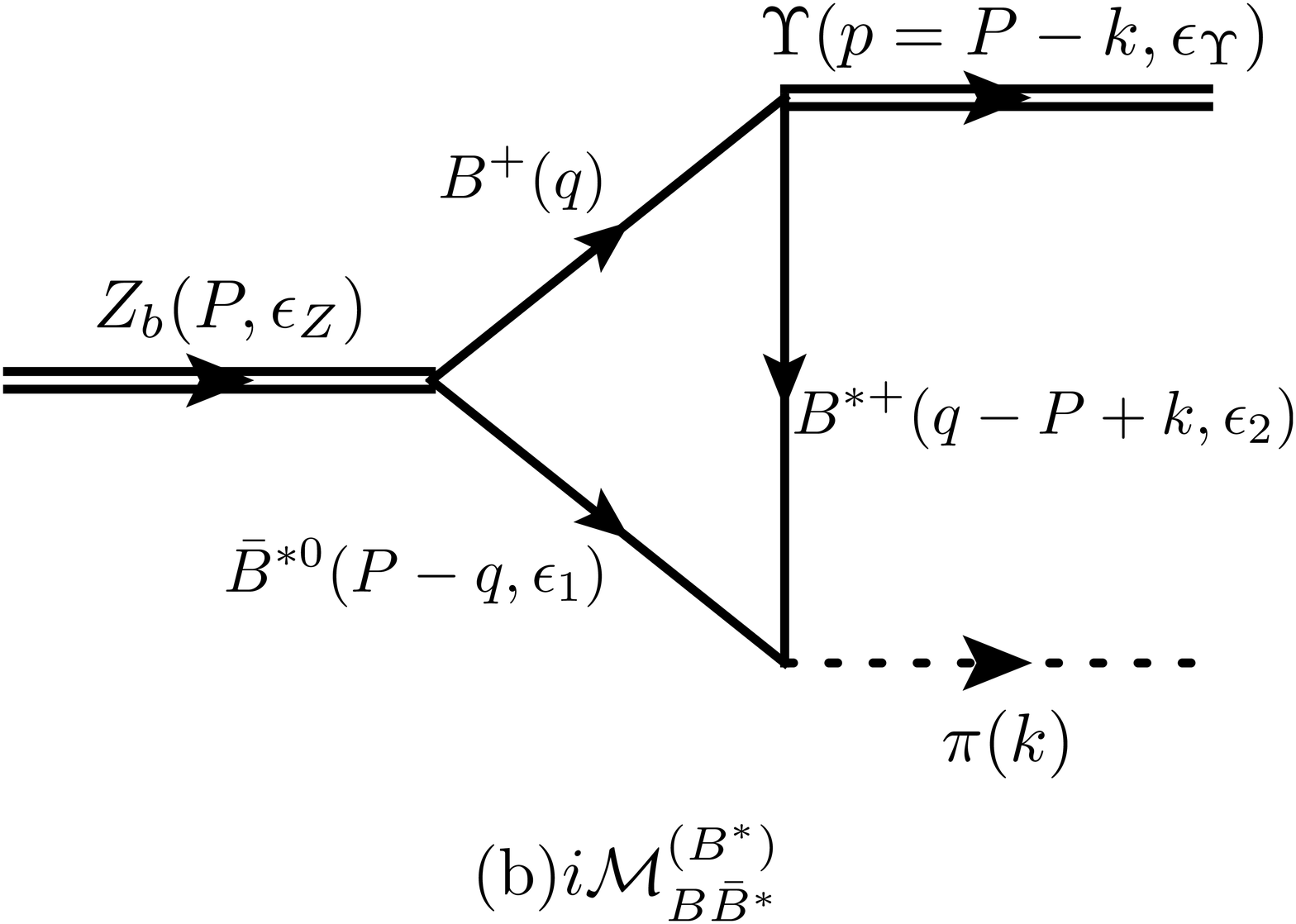} }\
 \subfigure{\
 \includegraphics[clip,width=50mm]{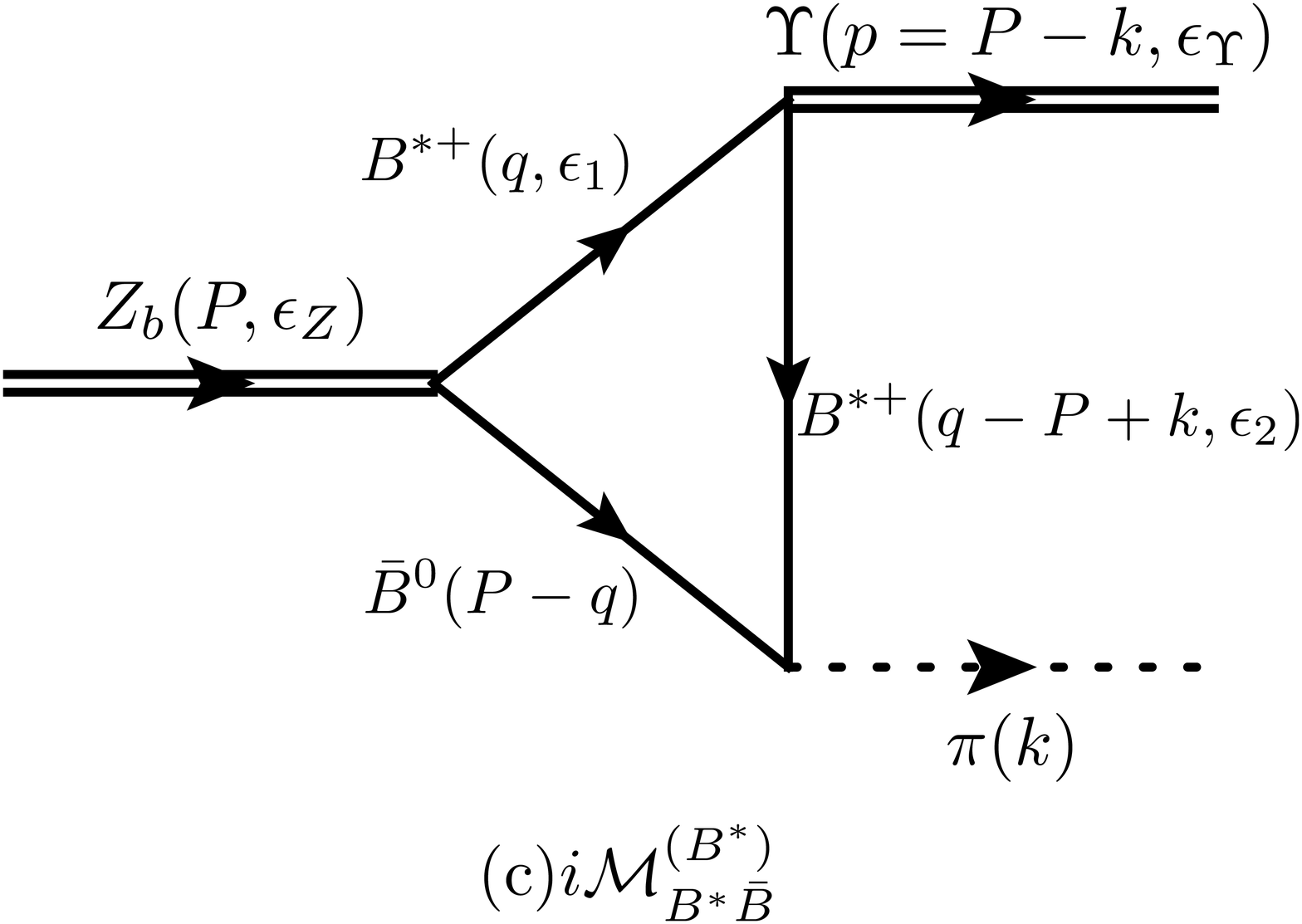} }\
 \caption{Feynman diagrams for $Z_b^+ \rightarrow \Upsilon(nS) \pi^+$.}
 \label{FDZb}
 \end{figure}
\begin{figure}[tbp]
 \subfigure{\ 
 \includegraphics[clip,width=50mm]{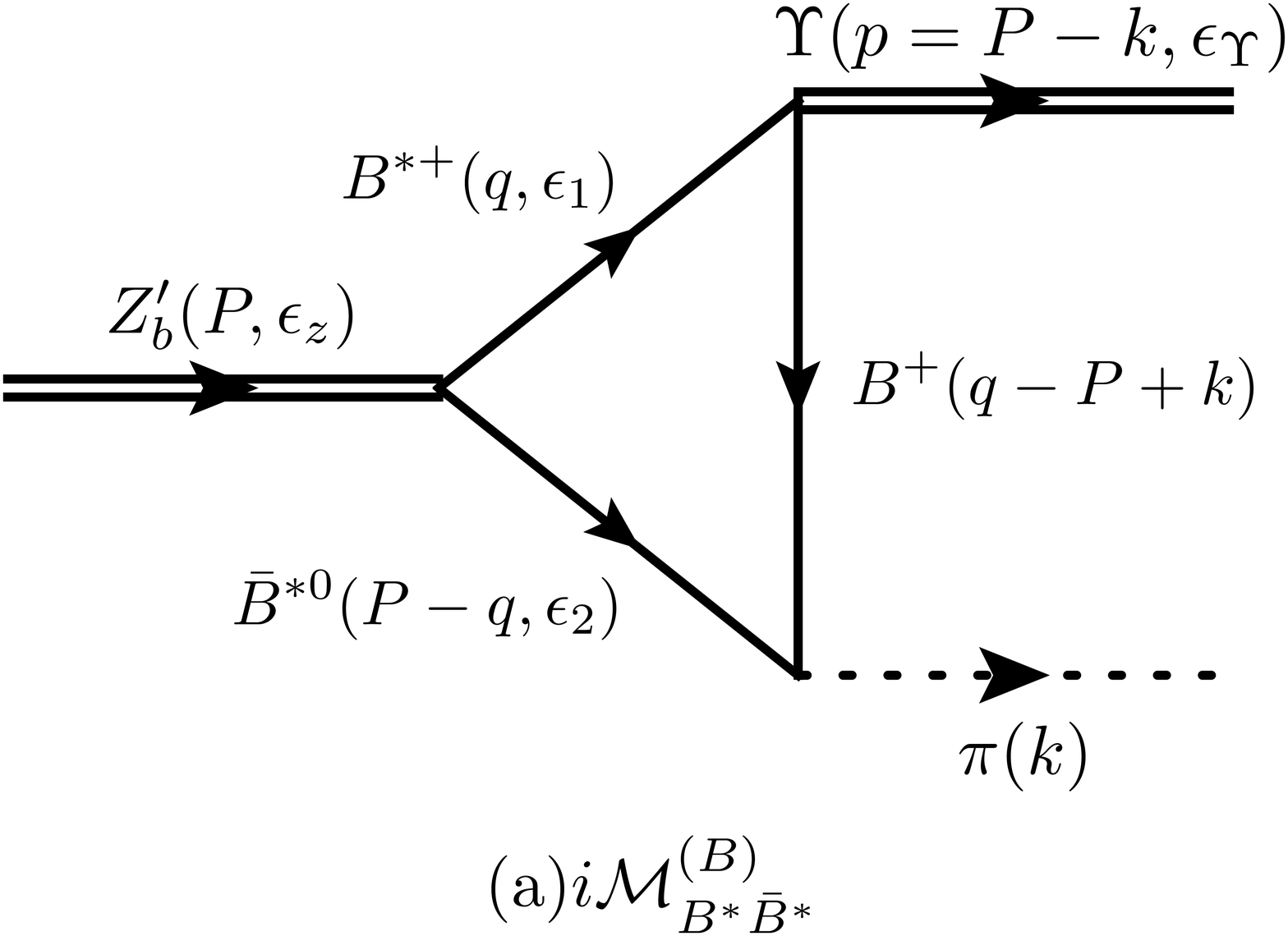} }\
 \subfigure{\
 \includegraphics[clip,width=50mm]{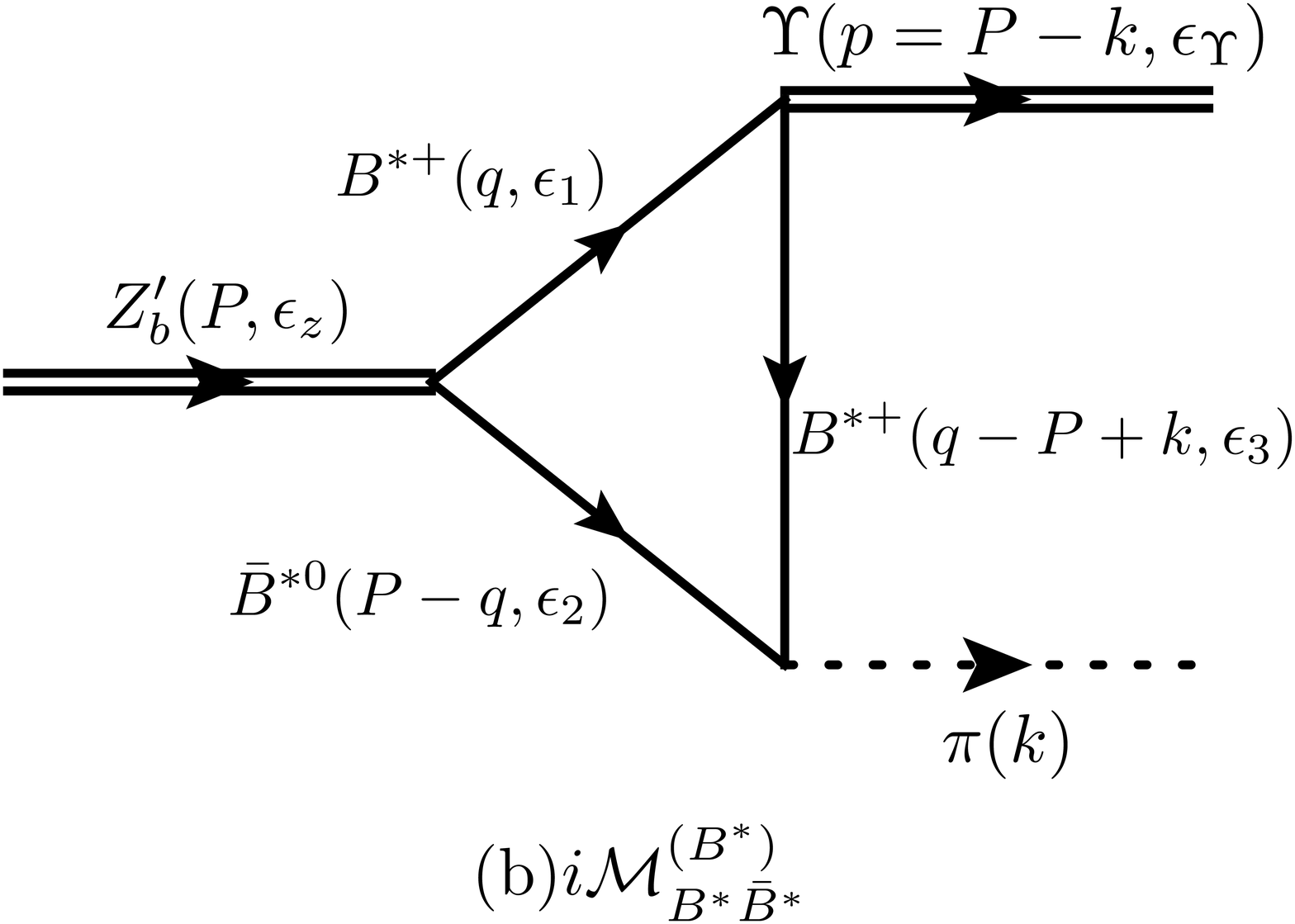} }\
 \caption{Feynman diagrams for $Z_b^{\prime +} \rightarrow \Upsilon(nS) \pi^+$.}
 \label{FDZbp}
 \end{figure}
%
To start the discussion,
we assume that the main components of $Z_b$ and $Z_b^{\prime}$ are molecular states of 
$\frac{1}{\sqrt{2}}(B\bar{B}^{\ast}-B^{\ast}\bar{B})(^3S_1)$ and 
 $B^{\ast}\bar{B}^{\ast}(^3S_1)$, namely,
\begin{eqnarray}
 \ket{Z_b} &=& \frac{1}{\sqrt{2}} \ket{B\bar{B}^{\ast} -B^{\ast}\bar{B}} \,, \\
 \ket{Z_b^{\prime}} &=& \ket{B^{\ast}\bar{B}^{\ast}} \,.
\end{eqnarray}
Such a simple molecular picture will give a good description, because those masses are close to the $B\bar{B}^{\ast}$ (or $B^{\ast}\bar{B}$) and $B^{\ast}\bar{B}^{\ast}$ thresholds,
respectively, and the ratio of $D$-wave mixing is not large.
In fact, the explicit calculations based on the hadronic model in our previous study indicate that the 
probability of the $\frac{1}{\sqrt{2}}(B\bar{B}^{\ast}-B^{\ast}\bar{B})(^3D_1)$ component is approximately 9 $\%$
and the $B^{\ast}\bar{B}^{\ast}(^3D_1)$ component is approximately 6 $\%$ in the total wave function of $Z_b$~\cite{Ohkoda:2011vj}.
In the hadronic molecular picture, the diagrams contributing to the decay $Z_b^{(\prime) +} \rightarrow 
\Upsilon (nS)\pi^+$ are described with the intermediate $B^{(\ast)}$ and $\bar{B}^{(\ast)}$ meson loops at lowest order~\cite{Cleven:2013sq,Li:2012as} as shown in
Figs.~\ref{FDZb} and \ref{FDZbp}.
Since $B^+$ and $\bar{B}^0$ are interchangeable, the total transition amplitudes are given by the twice of the sum of 
each channel as follows,
\begin{eqnarray}
 {\cal M}_{Z_b} &=& 2 ({\cal M}_{B \bar{B}^{\ast}}^{(B)} +{\cal M}_{B\bar{B}^{\ast}}^{(B^{\ast})} 
  +{\cal M}_{B^{\ast}\bar{B}}^{(B^{\ast})} )\,, \label{eq:amplitude1} \\
 {\cal M}_{Z_b^{\prime}} &=& 2 ({\cal M}_{B^{\ast}\bar{B}^{\ast}}^{(B)} 
  + {\cal M}_{B^{\ast}\bar{B}^{\ast}}^{(B^{\ast})} ) \,. \label{eq:amplitude2}
\end{eqnarray}

To calculate the transition amplitudes, we need the couplings from the effective Lagrangians.
We adopt the phenomenological Lagrangians at vertices of $Z_b^{(\prime)}$ and $B^{(\ast)}$ mesons,
which are
\begin{eqnarray}
 {\cal L}_{ZBB^{\ast}} &=& g_{ZBB^{\ast}} M_z Z^{\mu} (B B^{\ast \dagger}_{\mu} +B^{\ast}_{\mu}B^{\dagger}) 
  \,, \\
 {\cal L}_{Z^{\prime}B^{\ast}B^{\ast}} &=& i g_{Z^{\prime}B^{\ast}B^{\ast}} 
 \epsilon^{\mu \nu \alpha \beta} \partial_{\mu} Z_{\nu}^{\prime} B^{\ast}_{\alpha}
 B^{\ast \dagger}_{\beta} \,, 
\end{eqnarray}
where the coupling constants $g_{ZBB^{\ast}}$ and $g_{Z^{\prime}B^{\ast}B^{\ast}}$ are determined 
from the experimentally observed decay widths for the process to open heavy flavor channels from $Z_b^{(\prime)}$.
The experimental results are $\Gamma (Z_b^{+} \rightarrow 
B^{+}\bar{B}^{\ast 0} + B^{\ast +}\bar{B}^{0}) = 15.82 \,{\rm MeV}$ and 
$\Gamma (Z_{b}^{\prime +} \rightarrow B^{\ast +}\bar{B}^{\ast 0}) = 8.44 \,$MeV.
We obtain $g_{BB^{\ast}Z_b}= 1.04$ and $g_{B^{\ast}B^{\ast}Z_b^{\prime}}=1.30$ to reproduce
the observed values. 

For the other vertices, we employ the effective Lagrangians  reflecting both heavy quark symmetry and chiral 
symmetry~\cite{Colangelo:2003sa},
\begin{eqnarray}
 {\cal L}_{BB^{\ast}\pi} &=& -i g_{BB^{\ast}\pi} (B_i \partial_{\mu}\pi_{ij}B_j^{\dagger \ast \mu} 
 -B_i^{\ast \mu}\partial_{\mu}\pi_{ij}B_j^{\dagger}) \,, \\
 {\cal L}_{B^{\ast}B^{\ast}\pi} &=& \frac{1}{2}g_{B^{\ast}B^{\ast}\pi}\epsilon^{\mu\nu\alpha\beta}
 B^{\ast}_{i\mu}\overleftrightarrow{\partial}_{\alpha} \bar{B}^{\ast}_{j \beta} \partial_{\nu}
  \pi_{ij} \,, \\
  {\cal L}_{BB \Upsilon} &=& i g_{BB\Upsilon} \Upsilon_{\mu} (\partial^{\mu}BB^{\dagger} -B \partial^{\mu}B^{\dagger}) \,, \\
 {\cal L}_{BB^{\ast} \Upsilon} &=& -g_{BB^{\ast}\Upsilon} \epsilon^{\mu\nu\alpha\beta}
  \partial_{\mu}\Upsilon_{\nu} (\partial_{\alpha}B^{\ast}_{\beta}B^{\dagger} +B\partial_{\alpha}B^{\ast \dagger}_{\beta}) \,,  \\
 {\cal L}_{B^{\ast}B^{\ast}\Upsilon} &=& -i g_{B^{\ast}B^{\ast}\Upsilon} \left\{ \Upsilon^{\mu}
 (\partial_{\mu}B^{\ast \nu} B^{\ast \dagger}_{\nu} - B^{\ast \nu}\partial_{\mu}B^{\ast \dagger}_{\nu}) 
 +(\partial_{\mu}\Upsilon_{\nu}B^{\ast \nu} -\Upsilon_{\nu} \partial_{\mu}B^{\ast \nu})B^{\ast \dagger \nu}
 \right. \notag \\
 && \left. + B^{\ast \mu} (\Upsilon^{\nu}\partial_{\mu}B^{\ast \dagger}_{\nu}-\partial_{\mu}
 \Upsilon^{\nu}B^{\ast \dagger \nu} )\right\} \,,
\end{eqnarray}
where $B^{(\ast)}=(B^{(\ast)0},B^{(\ast)+})$.
The two coupling constants $g_{BB^{\ast}\pi}$ and $g_{B^{\ast}B^{\ast}\pi}$  are expressed by a  single 
 parameter $g$ thanks to heavy quark symmetry as follows:
\begin{align}
 g_{BB^{\ast}\pi} &= \frac{2g}{f_{\pi}}\sqrt{m_B m_{B^{\ast}}}\, , &
 g_{B^{\ast}B^{\ast}\pi} &= \frac{g_{BB^{\ast}\pi}}{\sqrt{m_B m_{B^{\ast}}}} \,, &
\end{align}
where $f_{\pi}=132$ MeV is a pion decay constant.
Since the decay $B^{\ast} \rightarrow B \pi$ is kinematically forbidden, it is impossible to determine the coupling $g$ from experiments.
Therefore, using the experimental information in the charm sector and the heavy quark symmetry, 
we adopt approximately $g=0.59$ when the observed decay width $\Gamma = 96$ keV for 
$D^{\ast} \rightarrow D\pi$ is used.
The coupling $g_{BB\Upsilon(nS)}$ of $\Upsilon(nS)$ and $B$ is estimated on the 
assumption of vector meson dominance (VMD)~\cite{Dong:2012hc}.
VMD gives the coupling constant 
$g_{BB\Upsilon(nS)} = M_{\Upsilon(nS)}/f_{\Upsilon(nS)}$, where $f_{\Upsilon(nS)}$ is a leptonic 
decay constant defined by 
$\braket{0|\bar{b}\gamma^{\mu}b| \Upsilon(nS)(p,\epsilon)} = f_{\Upsilon(nS)}\epsilon^{\mu}$.
Here $f_{\Upsilon(nS)}$ is determined from the leptonic decays $\Upsilon(nS) \rightarrow e^+ e^-$ as $f_{\Upsilon(1S)}=715$ MeV, $f_{\Upsilon(2S)}=497.5$ MeV and 
$f_{\Upsilon(3S)} = 430.2$ MeV, where the masses and decay widths are taken from Particle Data Group (PDG)~\cite{Beringer:1900zz}.
Thus we obtain $g_{BB\Upsilon(1S)}=13.2$, $g_{BB\Upsilon(2S)}=20.1$ and $g_{BB\Upsilon(3S)}=24.7$.
The other couplings $g_{BB^{\ast}\Upsilon(nS)}$ and $g_{B^{\ast}B^{\ast}\Upsilon(nS)}$ are related with 
$g_{BB\Upsilon(nS)}$ as
\begin{align}
 \frac{g_{BB\Upsilon(nS)}}{M_B} &= \frac{g_{BB^{\ast}\Upsilon(nS)}}{\sqrt{M_B M_{B^{\ast}}}} =
  -\frac{g_{B^{\ast}B^{\ast}\Upsilon(nS)}}{M_{B^{\ast}}} \,.
\end{align}
All the above arguments are valid in the heavy quark mass limit.
We neglect $1/m_Q$ corrections assuming that the mass of the bottom quark is sufficiently heavy.

In terms of the effective Lagrangians, we derive explicitly the transition amplitudes for 
$Z_b^{(\prime)} \rightarrow \Upsilon(nS) + \pi^+$ as follows:
\begin{eqnarray}
i{\cal M}^{(B)}_{BB^{\ast}} = &&
 (i)^3 \int \frac{d^4 q}{(2 \pi)^4} [ig_{ZBB^{\ast}} M_Z \epsilon_Z \cdot \epsilon_1]
[g_{BB\Upsilon(nS)}\left( \epsilon_{\Upsilon} \cdot (2q-p) \right)] 
[g_{B^{\ast}B^{\ast}\pi}(\epsilon_1 \cdot k)] \notag \\
&\times& \frac{1}{(q)^2 - m^2_{B}} \frac{1}{(P-q)^2 - m^2_{B^{\ast}}}
\frac{1}{(q-p)^2 - m^2_{B}} {\cal F}(\vec{q}^{\,\,2}, \vec{k}^{\,2}) \,, \\
 i {\cal M}_{BB^{\ast}}^{(B^{\ast})} = &&
 (i)^3  \int \frac{d^4 q}{(2 \pi)^4} [ig_{ZBB^{\ast}} M_Z \epsilon_Z \cdot \epsilon_1]
[g_{BB^{\ast}\Upsilon(nS)} i\epsilon_{\alpha \beta \gamma \delta} v^{\alpha}
\epsilon_{\Upsilon}^{\beta}\epsilon^{\gamma}_2 (2q-p)^{\delta}] \notag \\
&\times& [i\epsilon_{a b c d}g_{B^{\ast}B^{\ast}\pi}M_{B^{\ast}}v^{a}\epsilon_2^{b}k^c \epsilon_1^d]
\notag \\
&\times& 
\frac{1}{(q)^2 - m^2_{B}} \frac{1}{(P-q)^2 - m^2_{B^{\ast}}}
\frac{1}{(q-p)^2 - m^2_{B^{\ast}}} {\cal F}(\vec{q}^{\,\,2}, \vec{k}^{\,2}) \,, \\
 i{\cal M}_{B^{\ast}B}^{(B^{\ast})}= &&
 (i)^3 \int \frac{d^4 q}{(2 \pi)^4} [ig_{ZBB^{\ast}} M_Z \epsilon_Z \cdot \epsilon_1]
 [g_{BB^{\ast}\pi}] \notag \\
 &\times& \left[ g_{B^{\ast}B^{\ast}\Upsilon(nS)} \left\{(\epsilon_{\Upsilon}\cdot \epsilon_2) 
 \left( \epsilon_1 \cdot (2q-p)\right) +(\epsilon_{\Upsilon} \cdot \epsilon_1)
 \left( \epsilon_2 \cdot (2q-p)\right) -(\epsilon_1 \cdot \epsilon_2) \left( \epsilon_{\Upsilon}
 \cdot (2q-p)\right) \right\}\right] \notag \\
 &\times& 
 \frac{1}{(q)^2 - m^2_{B^{\ast}}} \frac{1}{(P-q)^2 - m^2_{B}}
 \frac{1}{(q-p)^2 - m^2_{B^{\ast}}} {\cal F}(\vec{q}^{\,\,2}, \vec{k}^{\,2}) \,,
\end{eqnarray}
\begin{eqnarray}
 i {\cal M}^{( B)}_{ B^{\ast}B^{\ast}} = && 
(i)^3 \int \frac{d^4 q}{(2 \pi)^4} [i g_{Z^{\prime}B^{\ast}B^{\ast}} 
\epsilon_{\mu \nu \alpha \beta} P^{\mu} 
\epsilon_z^{\nu} \epsilon_{1}^{\alpha} \epsilon_{2}^{\beta} ] \notag \\
&\times &  [ ig_{B^{\ast} B^{\ast} \Upsilon (nS)} \epsilon_{\delta \tau \theta \phi} v^{\delta} 
\epsilon_{\Upsilon}^{\tau} \epsilon_{1}^{\theta} (2q-p)^{\phi} ]
[g_{B B^{\ast}\pi} (\epsilon_{2} \cdot k) ] \notag \\ 
&\times &  \frac{1}{(q)^2 - m^2_{B^{\ast}}} \frac{1}{(P-q)^2 - m^2_{B^{\ast}}}
\frac{1}{(q-p)^2 - m^2_{B}} {\cal F}(\vec{q}^{\,\,2}, \vec{k}^{\,2}) \,, \\
  i {\cal M}^{( B^{\ast})}_{ B^{\ast}B^{\ast}} = && 
(i)^3 \int \frac{d^4 q}{(2 \pi)^4} [i g_{Z^{\prime}B^{\ast}B^{\ast}} 
\epsilon_{\mu \nu \alpha \beta} P^{\mu} 
\epsilon_z^{\nu} \epsilon_{1}^{\alpha} \epsilon_{2}^{\beta} ] 
[ig_{B^{\ast}B^{\ast}\pi} \epsilon_{0\tau\theta\phi}M_{B^{\ast}}\epsilon^{\tau}_{3}
k^{\theta}\epsilon_2] \notag \\
 &\times& \left[ g_{B^{\ast}B^{\ast}\Upsilon(nS)} \left\{(\epsilon_{\Upsilon}\cdot \epsilon_1) 
 \left( \epsilon_3 \cdot (2q-p)\right) +(\epsilon_{\Upsilon} \cdot \epsilon_3)
 \left( \epsilon_1 \cdot (2q-p)\right) -(\epsilon_1 \cdot \epsilon_3) \left( \epsilon_{\Upsilon}
 \cdot (2q-p)\right) \right\}\right] \notag \\
 &\times& \frac{1}{(q)^2 - m^2_{B^{\ast}}} \frac{1}{(P-q)^2 - m^2_{B^{\ast}}}
\frac{1}{(q-p)^2 - m^2_{B^{\ast}}} {\cal F}(\vec{q}^{\,\,2}, \vec{k}^{\,2}) \,,
\end{eqnarray}
where $P$ ($p$, $k$) is the momentum of $Z_{b}^{(')}$ ($\Upsilon(nS)$, $\pi$ meson), and $q$ is the momentum in the loop integrals.
We use the polarization vectors $\epsilon_{Z}$ and $\epsilon_{\Upsilon}$ for $Z_b^{(')}$ and $\Upsilon$ as well as $\epsilon_{1,2,3}$ for the propagating $B^{\ast}$ and $\bar{B}^{\ast}$ mesons in the loops.
To calculate the square of the absolute value of the transition amplitudes, we use the approximation for the polarization vector of the $B^{\ast}$ meson as 
$\epsilon_{B^{\ast}}^0 \simeq 0$ and use the sum over the polarizations $\lambda$ as 
$\sum_{\lambda}\epsilon^{\mu}_{B^{\ast}}\epsilon^{\nu}_{B^{\ast}} = \delta^{\mu\nu}$ ($\mu,\nu=1,2,3$) and 0 for other $\mu$ and $\nu$, because the absolute value of three-momentum $\vec{q}\,$ is assumed to be much smaller than the mass of $B^{(\ast)}$ meson in heavy quark limit~\cite{Aceti:2012cb}. 

In the above loop calculations, in order to reflect the finite range of the interaction, we use the form factor 
${\cal F}(\vec{q}^{\,\,2}, \vec{k}^{\,2})$ as follows,
\begin{eqnarray}
  {\cal F}(\vec{q}^{\,\,2}, \vec{k}^{\,2}) = 
   \frac{\Lambda_Z^2}{\vec{q}^{\,\,2}+\Lambda_Z^2}
   \frac{\Lambda^2}{\vec{k}^{\,2}+\Lambda^2}
   \frac{\Lambda^2}{\vec{k}^{\,2}+\Lambda^2 }\,.
\end{eqnarray}
The introduction of the form factor is important.
Since $Z_b^{(\prime)}$ is the loosely bound state of $B^{(\ast)}\bar{B}^{(\ast)}$,
the internal $B^{(\ast)}$ and $\bar{B}^{(\ast)}$ mesons move slowly almost as on-mass-shell particles.
According to this, the loop momentum $\vec{q}$ should be limited within a certain 
physical scale by the momentum cutoff parameter $\Lambda_Z$.
In a similar reason, the final state momentum $\vec{k}(=\vec{p}\,)$
will be controlled by a certain scale given by the momentum cutoff $\Lambda$ at vertices of $\Upsilon B^{(\ast)}B^{(\ast)}$ 
and $\pi B^{(\ast)}B^{(\ast)}$.
Thus, form factors with momentum cutoff are naturally introduced for each vertex from the view of the molecular picture.
Since the scale factors $\Lambda_Z$ and $\Lambda$ are related to the range of the hadron interaction,
they should be taken around the typical energy scale of hadron dynamics.
Thus, our formulation can include the finite range effects in a concise way and
regularize the amplitudes by the typical hadron scale.

\

We obtain the decay widths from the given amplitudes in Eqs.~(\ref{eq:amplitude1}) and (\ref{eq:amplitude2}).
As numerical inputs, all the masses are taken from the data of PDG~\cite{Beringer:1900zz}.
The numerical procedure is as follows: we integrate the amplitudes with $q^0$ analytically 
and pick up  poles in the propagators.
Since the masses of $Z_b^{(\prime)}$ are located above the $B\bar{B}^{\ast}$ (or $B^{\ast}\bar{B}$) and $B^{\ast}\bar{B}^{\ast}$ thresholds, respectively,
the integrals have singular points.
To treat them properly, we divide the integrals into real and imaginary parts by using 
the principle value of the integral.
In the end, it becomes possible to integrate with three-momentum $\vec{q}$ numerically.
This method can be naturally applied to the calculations of the amplitudes with the form factor. 
To confirm our calculations, we also adopt another method by a 
formalism of the Passarino-Veltman one-loop integral~\cite{Passarino:1978jh,Denner:1991kt}. 
We obtain an agreement in the numerical results between the two methods under the condition
of the large limit of scale factors ($\Lambda_Z$, $\Lambda \rightarrow \infty$). 

Tables~\ref{result:Zb} and \ref{result:Zbp} present the numerical results for the partial decay widths
 of $Z_b^{(\prime)}$. 
When the form factors are ignored, the decay widths are proportional to 
 $|\vec{k}|^{5}$, namely $\Gamma(Z_b^{(\prime)} \rightarrow \Upsilon(nS)\pi^+) \propto |\vec{k}|^{5}$.
This is much inconsistent with the experimental fact, 
because the loop integrals without form factors include the high-momentum contributions which are not acceptable in the low energy hadron dynamics.
In contrast, given the form factor,
 our calculations are qualitatively consistent with the experimental results:
(i) the decay to $\Upsilon(1S)\pi^+$ is strongly suppressed, (ii) the decay to $\Upsilon(2S)\pi^+$ occurs 
with the highest probability and (iii) the branching fraction of the decay to $\Upsilon(3S)\pi^+$ is smaller
than the one of $\Upsilon(2S)\pi^+$ but is still larger than the one of $\Upsilon(1S)\pi^+$.
We determine the cutoff parameters $\Lambda_Z = 1000$ MeV and 
$\Lambda = 600$ MeV 
 to reproduce the experimental values. To see the cutoff dependence, we change $\Lambda_Z$ as
$\Lambda_Z = 1000$, $1050$, $1100$ and $1150$ MeV and verified that the results do not change much.
The main reason for the suppression of the
$\Upsilon(1S)\pi^+$ decay is in the form factor depending on the 
final state momentum $\vec{k}$ ($\vec{p}\,$).
In contrast, this effect is minor for $\Upsilon(3S)\pi^+$ decay due to the small final state momentum.

\begin{table}[tbp]
\caption{The partial decay widths of $Z_b(10610)^+$ for various cutoff parameters $\Lambda_Z$ in units of  MeV. $\Lambda =600$ MeV is fixed. The left column shows the results without the form factors.}
\begin{tabular}{cccccc|c}
 $\Lambda_Z$ & - &$1000$& 1050 & $1100 $ &$1150$ & Exp.\\
 \hline
 $\Upsilon(1S) \pi^+$ & 96.3 & 0.074 & 0.079 & 0.083 & 0.087 &$0.059 \pm 0.017$ \\  
 $\Upsilon(2S) \pi^+$ & 20.0 & 0.47  & 0.50 & 0.52 & 0.55 & $0.81 \pm 0.22$\\
 $\Upsilon(3S) \pi^+$ & 0.498 & 0.14 & 0.14 & 0.15 & 0.15 & $0.40 \pm 0.10$  \\ 
\end{tabular} 
 \label{result:Zb}
\end{table}
\begin{table}[tbp]
\caption{The partial decay widths of $Z_b(10650)^+$. $\Lambda =600$ MeV is fixed. The unit is MeV.}
\begin{tabular}{cccccc|c}
 $\Lambda_Z$& - &$1000$& $1050$ & $1100 $ & $1150$ & Exp.\\
  \hline
 $\Upsilon(1S) \pi^+$ & 71.3 & 0.044 &0.046 & 0.049 & 0.051 &$0.028 \pm 0.008$ \\  
 $\Upsilon(2S) \pi^+$ & 17.6 & 0.31  & 0.33 & 0.34 & 0.36 & $0.28 \pm 0.07$\\
 $\Upsilon(3S) \pi^+$ & 0.858 & 0.18 & 0.19 & 0.20 & 0.21 & $0.19 \pm 0.05$  \\ 
\end{tabular} 
 \label{result:Zbp}
\end{table}

\begin{table}[tbp]
\caption{The partial decay widths of $Z_c^+$. $\Lambda =600$ MeV is fixed. The unit is MeV.}
\begin{tabular}{cccccc|c}
  $\Lambda_Z$ & - &$1000$& $1050$ & $1100 $ & $1150$ & Exp.\\
 \hline
 $J/\psi \pi^+$ & 39.0 & 0.66 & 0.69 & 0.71 & 0.73 & - \\  
 $\psi(2S) \pi^+$ & 0.305 & 0.18  & 0.17 & 0.17 & 0.18 & - \\
\end{tabular} 
 \label{result:Zc}
\end{table}
%

Finally, we briefly discuss the decays of $Z_c(3900)$ in the similar formalism, 
which has been recently observed in the 
$J/\psi \pi^+$ invariant mass spectrum of $Y(4260) \rightarrow J/\psi \pi^+\pi^-$ decay
by the BESIII Collaboration~\cite{Ablikim:2013mio}.
The reported mass and decay width are $M(Z_c) = 3899.0 \pm 3.6 \pm 4.9$ MeV and 
$\Gamma(Z_c) = 46 \pm 10 \pm 20$ MeV.
Belle collaboration also has reported $Z_c(3900)$ 
 with mass 
$M(Z_c) = 3894.5 \pm 6.6 \pm 4.5$ MeV and decay width $\Gamma(Z_c) = 63 \pm 24 \pm 26$ MeV~\cite{Liu:2013dau}.
Since $Z_c$ has the decay  properties and the mass spectrum both of which are similar to the $Z_b$ case,
it is expected that $Z_c$ would be the heavy-flavor partner of $Z_b$.
Thus, our model can apply to the analysis of the decays 
$Z_c \rightarrow J/\psi\pi^+$ and $\psi(2S)\pi^+$. 
In the present situation in experiments, branching fractions of $Z_c$ have not still been observed.
Besides, the decay $Z_c \rightarrow \psi(2S)$, which is allowed kinematically, is unconfirmed.
For these reasons, the numerical predictions are of benefit to the future experiments.

We apply the triangle diagram to the decays of $Z_c(3900)$.
We assume that $Z_c$ is a superposition state of $D\bar{D}^{\ast}$ and $D^{\ast}\bar{D}$, namely
\begin{eqnarray}
 \ket{Z_c} &=& \frac{1}{\sqrt{2}} \ket{D\bar{D}^{\ast} - D^{\ast}\bar{D}} \,.
\end{eqnarray}
The main difference between $Z_c^+ \rightarrow \psi(nS)\pi^+$ and 
$Z_b^+ \rightarrow \Upsilon(nS) \pi^+$ is the coupling constants for each vertex 
and masses of the hadrons.
As numerical inputs for $Z_c$, we use the averaged masses and decay widths reported by 
BESIII and Belle.
Considering that the branching fraction of $Z_b^{+} \rightarrow B^+ \bar{B}^{\ast 0} +B^{\ast +}\bar{B}^0$ 
is known to be 86.0 $\%$, we assume the one of 
$Z_c^{+} \rightarrow D^+ \bar{D}^{\ast 0} +D^{\ast +}\bar{D}^0$ is also approximately 86 $\%$ from the view of the heavy-flavor symmetry.
Then, we have the coupling $g_{Z_c DD^{\ast}}=2.23$ for $Z_{c}DD^{\ast}$ vertex.
The couplings $g_{DDJ/\psi} = 7.43$ and $g_{DD\psi(2S)}=12.4$ are employed with VMD.
Table~\ref{result:Zc} shows the numerical results for the partial decay widths of $Z_c$.
The width of $Z_c^+ \rightarrow \psi(2S)\pi^+$ is narrower than the one of 
$Z_c^+ \rightarrow J/\psi \pi^+$, owing to the small final state momentum.
The predicted branching fractions are $f(Z_c^+ \rightarrow J/\psi\pi^+) = 1.2 -1.3$ \% and 
$f(Z_c^+ \rightarrow \psi(2S) \pi^+) = 0.31 -0.33$ \%, 
which will be testable for future experiments.
Although $f(Z_c^+ \rightarrow \psi(2S) \pi^+)$ and $f(Z_b^+ \rightarrow \Upsilon(1S))\pi^+$ 
are almost same probabilities in our calculations, 
the main factors are different: the former is the narrow final phase space, the latter is 
the suppression due to the form factor.

\

In summary, we have studied the $Z_b^{(\prime)+} \rightarrow \Upsilon(nS) \pi^+$ decays
in a picture of the heavy meson molecule.
Assuming that $Z_b^{(\prime)}$ is the $B^{\ast}\bar{B}^{(\ast)}$ molecular state,
we have considered the transition amplitudes given by the triangle diagrams with $B^{(\ast)}$ and $\bar{B}^{(\ast)}$ meson loops at lowest order based on the heavy meson effective theory.
The couplings of $g_{ZBB^{\ast}}$ and $g_{Z^{\prime}B^{\ast}B^{\ast}}$  are fixed to reproduce correctly the observed decay widths
 from $Z_b^{(\prime)}$ to the open flavor channels.
To treat the effect of the finite range of the hadron interactions and 
regularize the loop integrals in the transition amplitudes suitably, 
we introduce the phenomenological form factors with 
the cutoff parameters $\Lambda_Z$ and $\Lambda$.
The numerical result with $\Lambda_z = 1000$ MeV and $\Lambda = 600$ MeV 
is qualitatively consistent with the experimental data.
Our results suggest that, if $Z_b^{(\prime)}$ have molecular type structures, 
the form factor should play a crucial role in the transition amplitudes.
Our model also applies the decays, $Z_c^+ \rightarrow J/\psi \pi^+ \,, \psi(2S)\pi^+$. 
We roughly estimate the branching fractions as 
$f(Z_c^+ \rightarrow J/\psi\pi^+) \sim 1.3$ \% and 
$f(Z_c^+ \rightarrow \psi(2S) \pi^+) \sim 0.32$ \%, 
which is testable for the future experiments in high energy accelerator facilities, such as KEK-Belle, BES and so on.
In the foreseeable future, our formulation will apply to the other exotic decays, such as
$Z_b^{(\prime)+} \rightarrow \eta_b \rho^+$, $Z_b^{(\prime)0} \rightarrow \eta_b \gamma$ and so on,
which can be also studied in future experiments.

\

 This work was supported  in part by Grant-in-Aid for JSPS Fellows (No.~15-5858 (S.~O.)) and 
 Scientific Research on Priority Areas
 ``Elucidation of New Hadrons with a Variety of Flavors (E01:21105006) (S.~Y. and A.~H.)"


\bibliographystyle{h-physrev5}
\bibliography{phys}

\end{document}